# Low Resolution Digital-to-Analog Converter with Digital Dithering for MIMO Transmitter

Arkady Molev-Shteiman, Xiao-Feng Qi, Laurence Mailaender

Radio Algorithms Research, Futurewei Technologies, Bridgewater, NJ, USA

*Abstract*— **Based on an equivalent model for quantizers with noisy inputs recently presented in [35], we propose a method of digital dithering at the transmitter that may significantly reduce the resolution requirements of MIMO downlink Digital to Analog Convertors (DAC). We use this equivalent model to analyze the effect of the dither Probability Density Function (PFD), and show that the uniform PDF produces an optimal (linear) result. Relative to other methods of DAC quantization error reduction our approach has the benefits of low computational complexity, compatibility with all existing standards, and blindness (no need for channel state information).**

## I. INTRODUCTION

Massive MIMO is an emerging technology capable of improving spectral efficiency of wireless communication by orders of magnitude. However a significant increase in base station antennas implies a proportional increase in cost and power consumption. On the other hand, it was shown that Massive MIMO may significantly mitigate the impact of imperfections in the hardware implementation [1],[2],implying that we may use cheaper and lower energy components to implement Massive MIMO. In the overall cost and energy budget of Massive MIMO base stations, quantizers, (ADCs and DACs) are important elements [3]. Therefore algorithms that reduce quantizer resolution have significant practical importance.

The problem of low-resolution AD/DA has received a lot of attention. Many contributions which consider uplink Massive MIMO receivers with arrays of low-resolution ADCs have been published [4]–[15]. However, attention has turned to the downlink transmitter only recently [16]–[27]. All these publications consider different methods of transmitter precoding that take low resolution DAC into account.

Here we propose digital dithering at the transmitter to mitigate the distortions caused by the low-resolution DAC. The use of dithering to reduce resolution of the ADC is well known. [27]-[30]. In [31] was shown that pseudo-noise injection improves the quality of images. The idea of using dither to improve the linearity of a DAC was proposed in [32] in 2017. However we believe that we are first to analyze the performance of the DAC in the context of a Massive MIMO system.

Our work is based on the equivalent model of a quantizer with noisy inputs that was recently presented in [35]. This model allows the analysis of dither PDFs that will minimize overall quantization error the DACs array.

It may turn out that dithering has lower performance than precoding, but this is compensated by following benefits:
- Low computation complexity.
- Generic solution for all modulation types
- No need for channel knowledge (blind).

In contrast to previous works that evaluated the low resolution DAC in terms of data throughput, we focus on the Error Vector Magnitude (EVM) of signals that arrive at the user. We believe that this criteria is most practically correct, because each communication standard determinates worst case distortion's level (maximal EVM) and if we don't want to change a standard (that is practically very challenging and time demanding) we have to meet standard requirement. Proposed method allows to reach this requirement by DAC with lower resolution.

## II. SYSTEM MODEL

We consider an 'all digital' Massive MIMO transmitter that has $M$ antennas equipped with $M$ pairs of DACs (one for real and one for imaginary signal). The goal of the transmitter is to deliver to each user $k$ desired signal $\tilde{y}_k$ with minimum distortion. We use a superscript '~' to denote a complex quantity. The total number of users is $K$. The desired signal $\tilde{y}_k$ could be any communication signal, for example OFDM.

Let us assume for simplicity that there are no multipath and channel is Line-Of-Sight only. Therefore, without taking into account the DAC resolution, a transmitter sends through each antenna $m$ a signal $\tilde{x}_m$ such that the sum signal that arrives at each user $k$ is equal to $\tilde{y}_k$:

$$\tilde{y}_k = \sum_{m=1}^{M} \tilde{c}_k(m) \cdot \tilde{x}_m \qquad (0.1)$$

Where $\tilde{c}_k(m)$ is the steering coefficient of the user $k$ at the antenna $m$. It satisfies $|\tilde{c}_k(m)| = 1$.

The actual realization of steering function depends on an antenna array configuration. For example, the steering function of uniform linear antenna array is given by:

$$\tilde{c}_k(m) = \exp\left(\sqrt{-1} \cdot \pi \cdot m \cdot \sin(\alpha_k)\right) \qquad (0.2)$$

where $\alpha_k$ is the direction to user $k$.

However, due to DAC quantization, actual signal that arrives to user $k$ is:

$$\hat{\tilde{y}}_k = \sum_{m=1}^{M} \tilde{c}_k(m) \cdot \tilde{Q}(\tilde{x}_m) \qquad (0.3)$$

Where complex quantization (DAC) operation defined as:
$$\tilde{Q}(\tilde{x}_m) = Q(\text{Re}(\tilde{x}_m)) + \sqrt{-1} \cdot Q(\text{Im}(\tilde{x}_m)) \quad (0.4)$$
where: $Q(x)$ denotes real quantization (DAC) operation.

For simplicity this paper does not consider the clipping effect of the DAC, we assume that the input signal is always within the dynamic range of the DAC,
$$-(2^{N-1}-1)\cdot\Delta \le x < (2^{N-1}-1)\cdot\Delta \quad (0.5)$$
where $\Delta$ and $N$ is the DAC quantization step and number of bits, respectively.

Within this band the quantization operation is given by:
$$Q(x) = \Delta \cdot round((x/\Delta)+0.5) - 0.5\cdot\Delta \quad (0.6)$$
where $round(\ )$ denote rounding operation. Throughput this paper, we assume a uniform quantizer is applied in the DAC.

For real signals the DAC output may be presented as the sum of desired signal and quantization error:
$$Q(x) = x + q \quad \text{where} \quad q = Q(x) - x \quad (0.7)$$

For sufficiently high DAC resolution we may approximate input signal PDF within each quantization step $\Delta$ as uniform. Then the PDF of the quantization error is also uniform within interval $\pm 0.5\cdot\Delta$. Therefore the quantization error has zero mean and variance equal to:
$$E[q^2] = \int_{q=-\infty}^{+\infty} q^2 \cdot p_q(q)\cdot dq = \int_{q=-0.5\Delta}^{+0.5\Delta} \frac{q^2}{\Delta}\cdot dx = \frac{\Delta^2}{12} \quad (0.8)$$
where $E[\ ]$ denotes random variable expectation.

The complex DAC output may be also presented as the sum of desired complex signal and complex quantization error:
$$\tilde{Q}(\tilde{x}_m) = \tilde{x}_m + \tilde{q}_m \quad \text{where} \quad \tilde{q}_m = \tilde{Q}(\tilde{x}_m) - \tilde{x}_m \quad (0.9)$$
This quantization error has zero mean and the variance:
$$E\left[|\tilde{q}_m|^2\right] = 2\cdot E[q^2] = \Delta^2/6 \quad (0.10)$$

From (0.1), (0.3) and (0.9) it follows that the actual signal that arrives at user $k$ is equal to:
$$\hat{\tilde{y}}_k = \tilde{y}_k + \sum_{m=1}^{M} \tilde{c}_k(m)\cdot \tilde{q}_m \quad (0.11)$$

The quantization distortion of signal arriving at user $k$ is measured by Error Vector Magnitude (EVM), here defined as the ratio between variance of distortion (quantization error) and variance of desired signal,
$$EVM_k = E\left[|\hat{\tilde{y}}_k - \tilde{y}_k|^2\right]/E\left[|\tilde{y}_k|^2\right] = E\left[\left|\sum_{m=1}^M \tilde{c}_k(m)\cdot\tilde{q}_m\right|^2\right]/E\left[|\tilde{y}_k|^2\right] \quad (0.12)$$

Each communication standard specifies the worst case distortion level (or, maximal EVM) of the signal that arrives at a user. Based on this, we can determine the minimal DAC resolution that generates distortions below this limit.

From (0.12) it follows that the signal EVM is minimal when there is any correlation between different quantization error realizations. When this happens, according to (0.10):
$$\min(EVM_k) = \sum_{m=1}^{M}|\tilde{c}_k(m)|^2\cdot E\left[|\tilde{q}_m|^2\right]/E\left[|\tilde{y}_k|^2\right] = \frac{M\cdot\Delta^2/6}{E\left[|\tilde{y}_k|^2\right]} \quad (0.13)$$

The EVM is maximal (worst case) when different quantization error realizations are fully correlated and satisfy:
$$\tilde{q}_m = \left(\tilde{c}_k(m)/\tilde{c}_k(1)\right)^*\cdot \tilde{q}_1 \quad (0.14)$$

where $(\ )^*$ denotes conjugate operation.

When this happens, the quantization errors of all antennas sum coherently and therefore according to (0.10):
$$\max(EVM_k) = \left|\sum_{m=1}^{M}\tilde{c}_k(m)\right|^2\cdot E\left[|\tilde{q}_m|^2\right]/E\left[|\tilde{y}_k|^2\right] = \frac{M^2\cdot\Delta^2/6}{E\left[|\tilde{y}_k|^2\right]} \quad (0.15)$$

An example of such a worst case is the uniform linear array, when a single user arrives from direction equal to:
$$\alpha_1 = 0, \pm\pi/2, \pm\pi/3 \quad (0.16)$$
According to (0.2) for such $\alpha_1$
$$\tilde{c}_1(m) = (\pm 1 \text{ or } \pm\sqrt{-1}) \quad (0.17)$$
In single user scenario each antenna $m$ sends the signal:
$$\tilde{x}_m = \tilde{c}_1(m)^*\cdot \tilde{y}_1/M \quad (0.18)$$
According to (0.4) and (0.6)
$$\tilde{Q}(-\tilde{x}) = -\tilde{Q}(\tilde{x}) \quad \text{and} \quad \tilde{Q}(\sqrt{-1}\cdot\tilde{x}) = \sqrt{-1}\cdot\tilde{Q}(\tilde{x}) \quad (0.19)$$
and according to (0.9):
$$\tilde{q}_m = \tilde{Q}(\tilde{x}_m) - \tilde{x}_m = \tilde{Q}\left(\tilde{c}_1(m)^*\cdot\tilde{y}_1/M\right) - \left(\tilde{c}_1(m)^*\cdot\tilde{y}_1/M\right) =$$
$$= \tilde{c}_1(m)^*\cdot\left(\tilde{Q}(\tilde{y}_1/M) - (\tilde{y}_1/M)\right) = \left(\tilde{c}_k(m)/\tilde{c}_k(1)\right)^*\cdot\tilde{q}_1 \quad (0.20)$$
We may see that the quantization error (0.20) satisfies the worst case scenario definition of (0.14), hence the DAC resolution must be set according to (0.15).

III. EQUIVALENT MODEL OF QUANTIZER WITH NOISY INPUT

Here we apply the quantizer equivalent model [35], originally developed for ADCs, to the DAC with dithered input.

Let us consider an array of $M$ complex quantizers (complex DACs) where each quantizer input $m$ is the sum of the desired signal $\tilde{x}_m$ and an additive dither $\tilde{w}_m$. Assume that dither signals across quantizers are identical and independently distributed. Assume the real and imaginary parts of the dither are mutually independent and have the same PDF $p_W(\ )$.

Define the complex DAC equivalent transfer function as:
$$\tilde{F}(\tilde{x}) = E\left[\tilde{Q}(\tilde{x}+\tilde{w})|\tilde{x}\right] = F(\text{Re}(\tilde{x})) + \sqrt{-1}\cdot F(\text{Im}(\tilde{x})) \quad (0.21)$$
where the real quantizer equivalent transfer function is the expectation of the quantizer output given the quantizer desired input signal. It equals the time-reversed convolution of the original quantizer transfer function with the dither PDF:
$$F(x) = E[Q(x+w)|x] = \int_{x=-\infty}^{\infty} \tilde{Q}(x+w)\cdot p_W(w)\cdot dw \quad (0.22)$$
Let us define quantizer equivalent output noise as,
$$\tilde{n}_m = \tilde{Q}(\tilde{x}_m + \tilde{w}_m) - \tilde{F}(\tilde{x}_m) \quad (0.23)$$
Therefore we may represent output of such DAC as the sum of DAC desired input signal that passes through equivalent non-linear element and equivalent noise additive noise.
$$\tilde{Q}(\tilde{x}_m + \tilde{w}_m) = \tilde{F}(\tilde{x}_m) + \tilde{n}_m \quad (0.24)$$
In [35] the quantizer equivalent output noise was shown to have the following properties,
$$E[\tilde{n}_m] = 0 \quad \text{for any } m \quad (0.25)$$
$$E\left[\tilde{n}_{m1}^*\cdot\tilde{x}_{m2}\right] = 0 \quad \text{for any } m1 \text{ and } m2 \quad (0.26)$$
$$E\left[\tilde{n}_{m1}^*\cdot\tilde{n}_{m2}\right] = 0 \quad \text{for any } (m1 \ne m2) \quad (0.27)$$

## IV. DAC WITH OPTIMAL DITHER

Let's assume that each DAC has an independent digital additive dither having a uniform PDF over the interval $\pm \Delta/2$.

$$\text{if}\,(-\Delta/2 \leq w < +\Delta/2)\ p_W(w) = 1/\Delta\ \text{else}\ p_W(w) = 0 \quad (0.28)$$

Then according to (0.22) with DAC dynamic range(0.5), the equivalent transfer function of the real quantizer is,

$$F(x) = E[Q(x+w)|x] = x_L(x) \cdot \Pr(Q(x+w) = x_L(x)|x) + \\ + x_U(x) \cdot \Pr(Q(x+w) = x_U(x)|x) \quad (0.29)$$

where $x_L(x)$ and $x_U(x)$ are the two possible neighboring DAC outputs adjacent to DAC input value $x$. According to (0.6):

$$x_L(x) = \Delta \cdot fix((x+0.5\cdot\Delta)/\Delta) - 0.5\cdot\Delta \quad (0.30)$$

$$x_U(x) = \Delta \cdot fix((x+0.5\cdot\Delta)/\Delta) + 0.5\cdot\Delta = x_L + \Delta \quad (0.31)$$

where $fix(x)$ denotes the integer part of a real number $x$.

The probability of each DAC output given DAC input $\tilde{x}$ is equal to:

$$\Pr(Q(x+w) = x_L(x)|x) = \Pr(w \leq (x_L(x) + 0.5\cdot\Delta - x)) \\ = \int_{w=-0.5\cdot\Delta}^{x_L(x)+0.5\cdot\Delta - x} \frac{w}{\Delta} \cdot d\tilde{w} = \frac{x_L(x) + \Delta - x}{\Delta} = \frac{x_U(x) - x}{\Delta} \quad (0.32)$$

$$\Pr(Q(x+w) = x_U(x)|x) = \Pr(w > (x_L(x) + 0.5\cdot\Delta - x)) \\ = \int_{w=x_L(x)+0.5\cdot\Delta - x}^{0.5\cdot\Delta} \frac{w}{\Delta} \cdot dw = \frac{x - x_L(x)}{\Delta} \quad (0.33)$$

Therefore according to (0.29), (0.32) and (0.33) the expectation of the DAC output given the DAC input $x_I$ is equal to:

$$F(x) = (x_L(x)\cdot(x_U(x)-x) + x_U(x)\cdot(x-x_L(x)))/\Delta = x \quad (0.34)$$

Figure 1 illustrates the linearization effect of the convolution (0.22) on the equivalent DAC transfer function.

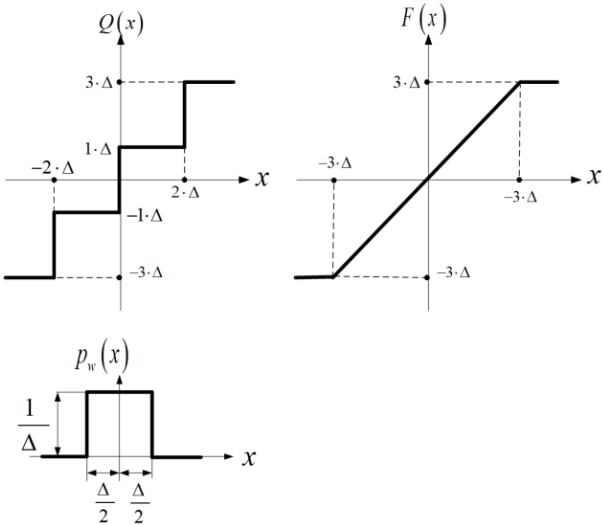

Figure 1    The 2 bit DAC equivalent transfer function

According (0.34) to (0.21) the complex quantizer equivalent transfer function with DAC dynamic range (0.5) is linear.

$$\tilde{F}(\tilde{x}) = \tilde{x} \quad (0.35)$$

Hence the uniform dither distribution (0.28) is optimal for the uniform DAC. From (0.35) and (0.24) it follows that we may represent the output of the quantizer with optimal dither as the sum of the desired signal and the equivalent additive noise,

$$\tilde{Q}(\tilde{x}_m + \tilde{w}_m) = \tilde{x}_m + \tilde{n}_m \quad (0.36)$$

In contrast to a conventional DAC's quantization error, the equivalent additive noise $\tilde{n}_m$ is a white process that satisfies (0.25), (0.26) and (0.27).

Let us first calculate the variance of the real and imaginary parts of additive noise $\tilde{n}_m$. According to (0.32) and (0.33) it is,

$$E[n^2] = E[\operatorname{Re}(\tilde{n}_m)^2] = E[\operatorname{Im}(\tilde{n}_m)^2] = E[(Q(x+w)-x)^2] = \\ = E[(x-x_L(x))^2 \cdot \Pr(Q(x+w) = x_L(x)|x)] + \\ + E[(x_U(x)-x)^2 \cdot \Pr(Q(x+w) = x_U(x)|x)] = \\ = E\left[(x-x_L(x))^2 \cdot \frac{x_U(x) - x}{\Delta} + (x_U(x)-x)^2 \cdot \frac{x - x_L(x)}{\Delta}\right] = \\ = E[(x-x_L(x))\cdot(x_U(x)-x)] = E[x_E \cdot (\Delta - x_E)] \quad (0.37)$$

where $x_E = x - x_L(x)$ satisfies to $0 \leq x_E < \Delta$

Let us again approximate the input signal PDF within each interval $\Delta$ as uniform. Then the PDF of $x_E$ is also uniform within interval $0 \leq x_E < \Delta$. Therefore we may express the real quantization error variance as,

$$E[n^2] = E[x_E \cdot (\Delta - x_E)] = \int_{x_E=0}^{\Delta} \frac{x_E \cdot (\Delta - x_E)}{\Delta} \cdot dx_E = \frac{\Delta^2}{6} \quad (0.38)$$

therefore, variance of the complex quantization error is:

$$E[|\tilde{n}_m|^2] = 2 \cdot E[n^2] = \Delta^2/3 \quad (0.39)$$

The resulting quantization noise variance of the DAC with dither is 3dB higher than that of a conventional DAC, which according to (0.10) is $\Delta^2/6$.

This 3dB degradation of noise power is the price we pay in order to make the quantization noises across all the complex DACs within the array uncorrelated with each other(0.27). The benefit of this becomes apparent in the next section.

## V. PERFORMANCE OF MIMO TRANSMITTER WITH OPTIMALLY DITHERED DACS

From (0.1), (0.3) and (0.36) the signal that arrives at user $k$ from the array of DACs with dither is,

$$\hat{\tilde{y}}_k = \tilde{y}_k + \sum_{m=1}^{M} \tilde{c}_k(m) \cdot \tilde{n}_m \quad (0.40)$$

As the quantization error of complex DACs with dither has variance $\Delta^2/3$ (0.39) and satisfies (0.27), the EVM (0.12) of the signal at user $k$ is,

$$EVM_k = \sum_{m=1}^{M} |\tilde{c}_k(m)|^2 \cdot E[|\tilde{n}_m|^2] \Big/ E[|\tilde{y}_k|^2] = \frac{M \cdot \Delta^2/3}{E[|\tilde{y}_k|^2]} \quad (0.41)$$

Thus we see its performance is only half as good as that of a conventional DAC in the best case of (0.13), but $(M/2)$ times as good as that of the conventional DACs in the worst case of (0.15) and the latter defines the necessary DAC resolution. In order to always satisfy the same EVM requirement, the array of *M* complex DACs with optimal dither can tolerate a quantization step that is $\sqrt{M/2}$ as large as that of the conventional DACs, with the same size-*M* array.

$$\Delta_D/\Delta_C = \sqrt{M/2} \quad (0.42)$$

Therefore for a fixed EVM requirement, the array of *M* DACs with optimal dither needs $\log_4(M/2)$ fewer bits than the same array with conventional DACs.

$$N_C/\Delta_D = \log_2(\Delta_D/\Delta_C) = \log_4(M/2) \quad ](0.43)$$

## VI. SIMULATION RESULTS

We illustrate the concept with the following numerical experiment.
- We simulate a single user, *M* antenna MIMO downlink transmitter. Each antenna *m* signal $x_m$ is given by(0.18). Each antenna is equipped with a pair of *N*-bits DACs. *N* and *M* are simulation parameters.
- We assume a uniform linear MIMO array with steering coefficients equal to(0.2). The user's direction $\alpha_1$ is a simulation parameter.
- We assume that the signal $y_1$ that we send to user has Gaussian distribution which is typical for OFDM modulation. In order to avoid clipping effect we set the ratio between input signal RMS and DAC maximal value to 15dB.

$$Peak2rms = (2^{N-1} \cdot \Delta)^2 / E[x_m^2] = 15\ dB \quad (0.44)$$

- We measure the EVM of the signal at the user for each direction $\alpha_1$ from -90 to +90 degree with a 3 degree step. Each simulation point is averaged over 1e4 samples.

Figure 2 presents the EVM of the signal arriving at user versus the angle of departure for an array of 1000 antenna elements, each equipped with either 6-bit conventional DACs or 6-bit DACs with dither. For reference we also present the analytical curve of the dithered DAC EVM calculated according to(0.1),(0.41) and (0.44)as:

$$EVM_k = \frac{M \cdot \Delta^2}{3 \cdot E[|\tilde{y}_k|^2]} = \frac{\Delta^2}{3 \cdot M \cdot E[x_m^2]} = \frac{Peak2rms}{3 \cdot M \cdot (2^{2\cdot(N-1)})} \quad (0.45)$$

Figure 3 presents the worst case (angle of departure is 0) EVM of signal arriving at the user from a MIMO array with 1,10,100,1000 and 10000 antenna elements each equipped with a pair of dithered DACs, as a function of DACs resolution. For reference we also present the performance of convention DAC that in the worst case is independent of the number of antenna elements.

From these figures we can make the following observations:

- The analytical EVM curve of DAC with dither fully matches with simulation results.
- The performance of the DAC with dither does not depend on the user's direction. However performance of conventional DAC does. The worst case user's direction for conventional DAC that satisfies to (0.16) .
- Even in a typical case, the performance of the DAC with dither is significantly better than the conventional one
- In the worst case scenario, the DAC with dither provides a gain equal to $10 \cdot \log_{10}(M/2)$ dB. For a single antenna transmitter, the DAC with dither causes 3dB performance degradation.

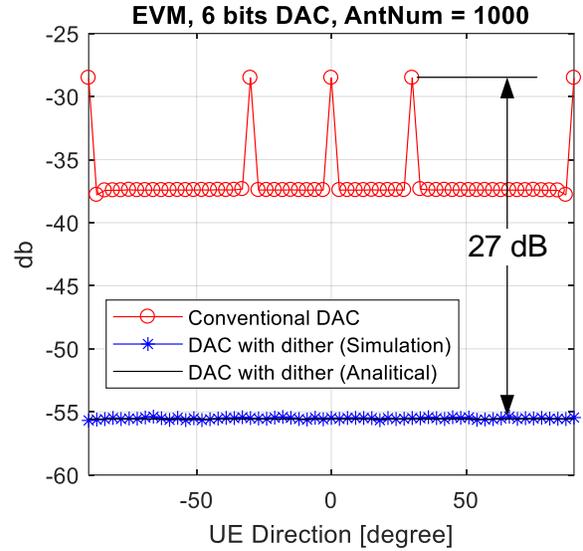

Figure 2  The EVM of signal arriving at user from MIMO downlink transmitter with array of 1000 antennas

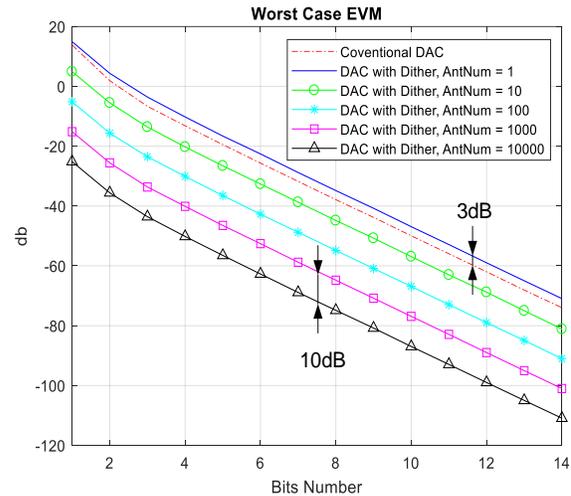

Figure 3  The worst case EVM of signal arriving at user as function of DAC resolution and antenna number

## VII. CONCLUSIONS.

We have shown that the DAC with optimal dither improves the EVM of signal arriving at a user from a size-*M* MIMO array by a factor proportional to the array size, given the same resolution. Alternatively, the same worst case EVM can be maintained with $\log_4(M/2)$ fewer bits. Furthermore, the EVM is now equal in all directions, in contrast to the case of an array equipped with conventional DACs.